\documentclass[a4paper]{article}

\usepackage{spconf,amsmath,graphicx,cite,amssymb}
\usepackage{amsthm,moresize}
\usepackage[tight,footnotesize]{subfigure}
\title{UNDERDETERMINED SOURCE SEPARATION USING A SPARSE STFT FRAMEWORK AND WEIGHTED LAPLACIAN DIRECTIONAL MODELLING}
\name{Thomas Sgouros and Nikolaos Mitianoudis}
\address{Electrical and Computer Engineering Department\\
Democritus University of Thrace\\
Xanthi, Greece}

\begin{document}

\maketitle
\begin{abstract}
The instantaneous underdetermined audio source separation problem  of $K$-sensors, $L$-sources mixing scenario (where $K<L$) has been addressed by many different approaches, provided the sources remain quite distinct in the virtual positioning space spanned by the sensors. This problem can be tackled as a directional clustering problem along the source position angles in the mixture. The use of Generalised Directional Laplacian Densities (DLD) in the MDCT domain for underdetermined source separation has been proposed before. Here, we derive weighted mixtures of DLDs in a sparser representation of the data in the STFT domain to perform separation. The proposed approach yields improved results compared to our previous offering and compares favourably with the state-of-the-art.
 \end{abstract}
\section{Introduction}
\label{sec:intro}
Assume a set of $K$ microphones $\mathbf{x}(n)=[x_1(n),\dots,x_K(n)]^T$ observing a set of $L$ $(K<L)$ sound sources $\mathbf{s}(n)=[s_1(n),\dots,s_L(n)]^T$. The instantaneous (anechoic) mixing model can  be expressed as follows:
\begin{equation}
\mathbf{x}(n)=\mathbf{A}\mathbf{s}(n)
\label{Mix}
\end{equation}
where $\mathbf{A}$ represents a $K\times L$ {\em mixing matrix} and $n$ the sample index of $N$ avaivable data samples. Blind Source Separation (BSS) algorithms provide an estimate of the source signals $\mathbf{s}$ and the mixing matrix $\mathbf{A}$, based on the observed microphone signals and some general statistical source profile. A variety of algorithms provide hiqh-quality separation solutions for the complete instantaneous case ($K=L$) \cite{ICAbook02}. The underdetermined instantaneous case is more challenging, since the estimation of the mixing matrix $\mathbf{A}$ alone is not sufficient to complete the separation~\cite{Mitianoudis07c}.

Many solutions exist for the underdetermined source separation problem. A good survey of underdetermined methods for source separation can be found in \cite{ICAbook02}. Recently, Arberet et al~\cite{Arberet10} proposed a method to count and locate sources in underdetermined mixtures. Their approach is based on the hypothesis that in localised neighbourhoods around time-frequency points $(t,f)$ (in the Short-Time Fourier Transform (STFT) representation) only one source essentially contributes to the mixture. Thus, they estimate the most dominant source and a local confidence measure, which shows where a single component is only present. A clustering approach merges the above information and estimates the mixing matrix $\mathbf{A}$. In~\cite{Vincent09}, Vincent et al used local Gaussian Modelling of minimal constrained variance of the local time-frequency neighbours assuming knowledge of the mixing matrix $\mathbf{A}$. The candidate sources' variances are estimated after minimising the Kullback-Leibler (KL) divergence between the empirical and expected mixture covariances, assuming that at most 3 sources contribute to each time-frequency neighbourhood  and the sources are derived using Wiener filtering. 
\begin{figure*}
\centering
\subfigure[Initial scatter plot]{\includegraphics[width=1.7in]{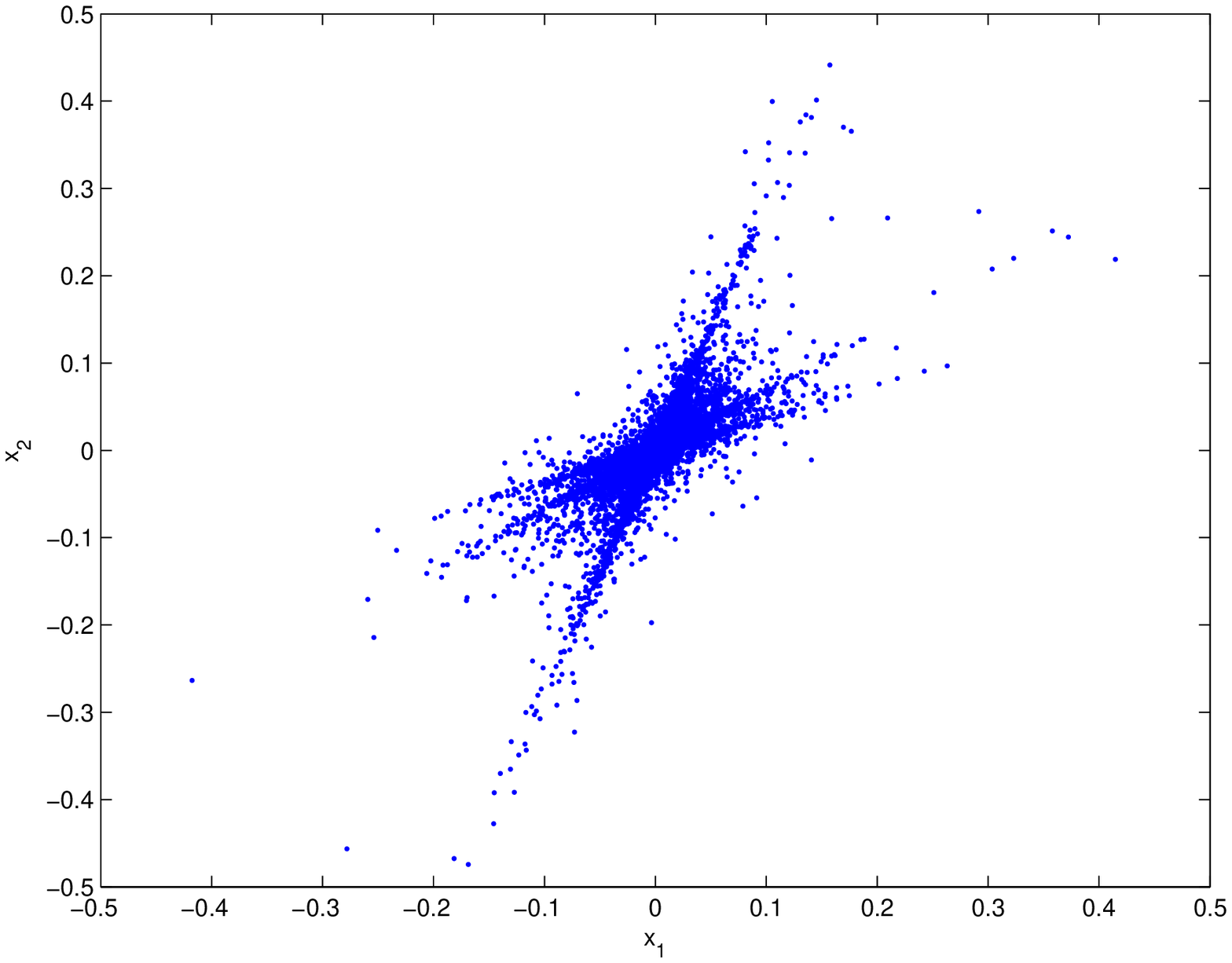}}
\subfigure[Sparsification using \cite{Mitianoudis07f} ]{\includegraphics[width=1.7in]{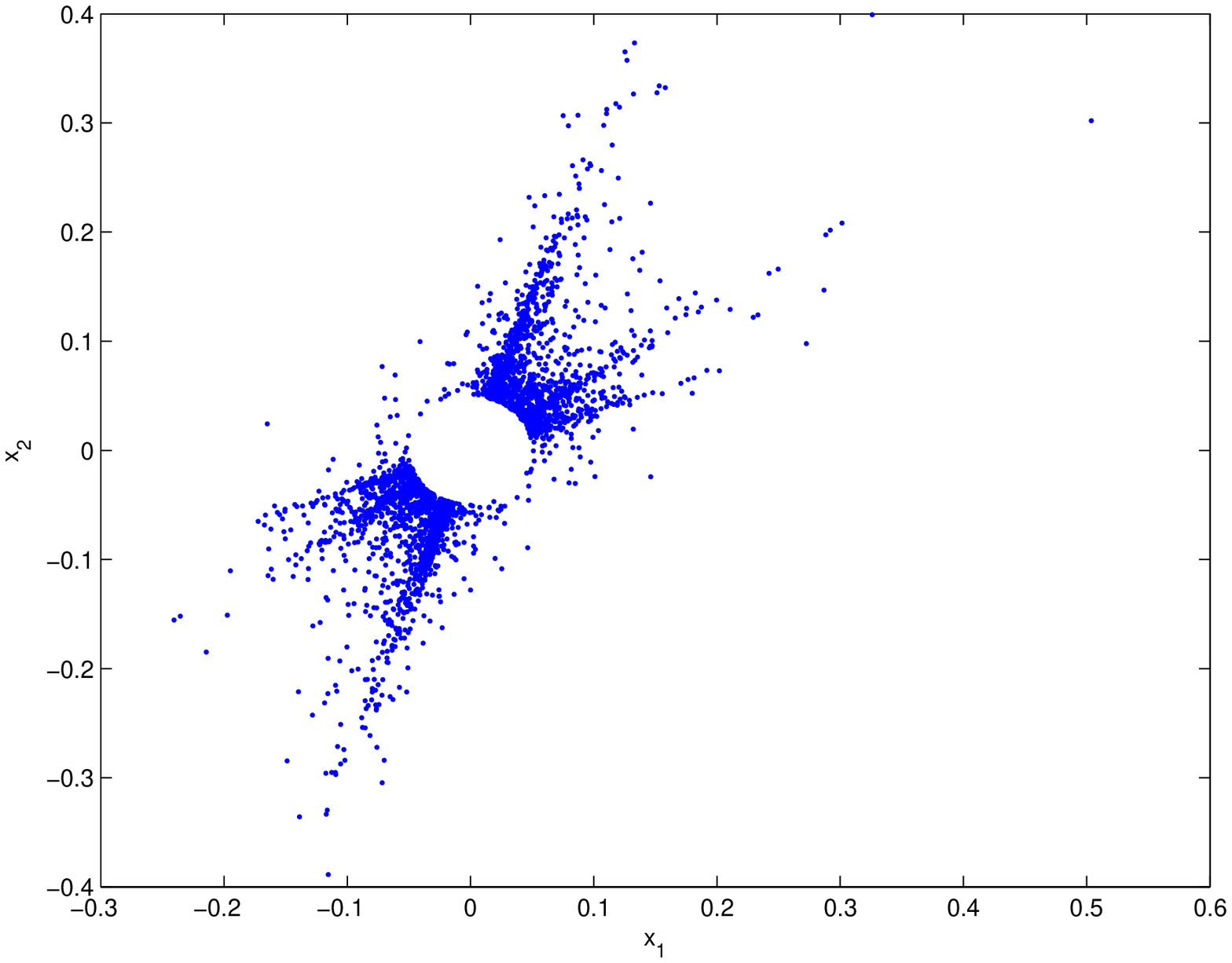}}
\subfigure[Sparsification using \cite{Arberet10} ]{\includegraphics[width=1.7in]{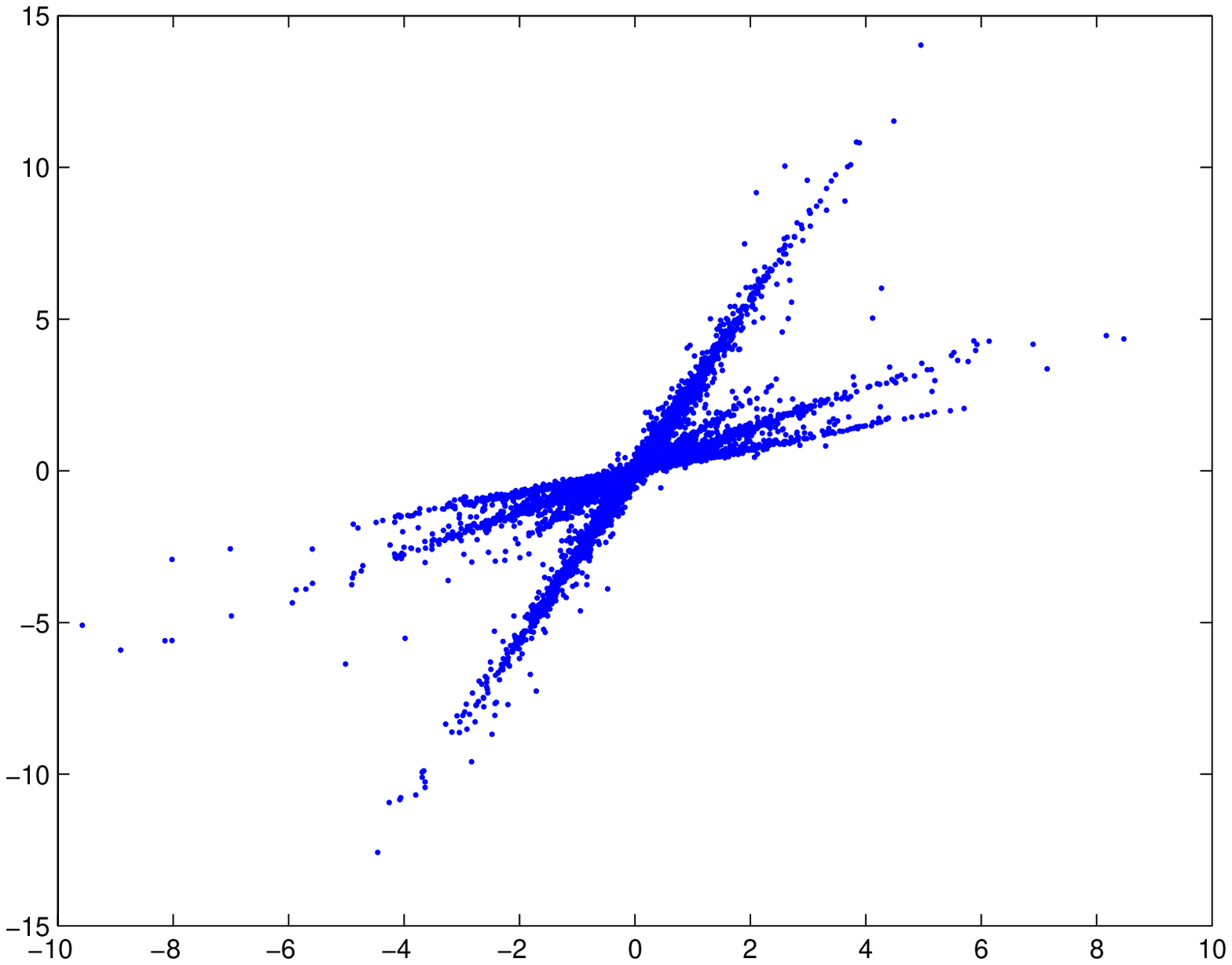}}\\
\subfigure[Intial Histogram of $\theta_n$]{\includegraphics[width=1.7in]{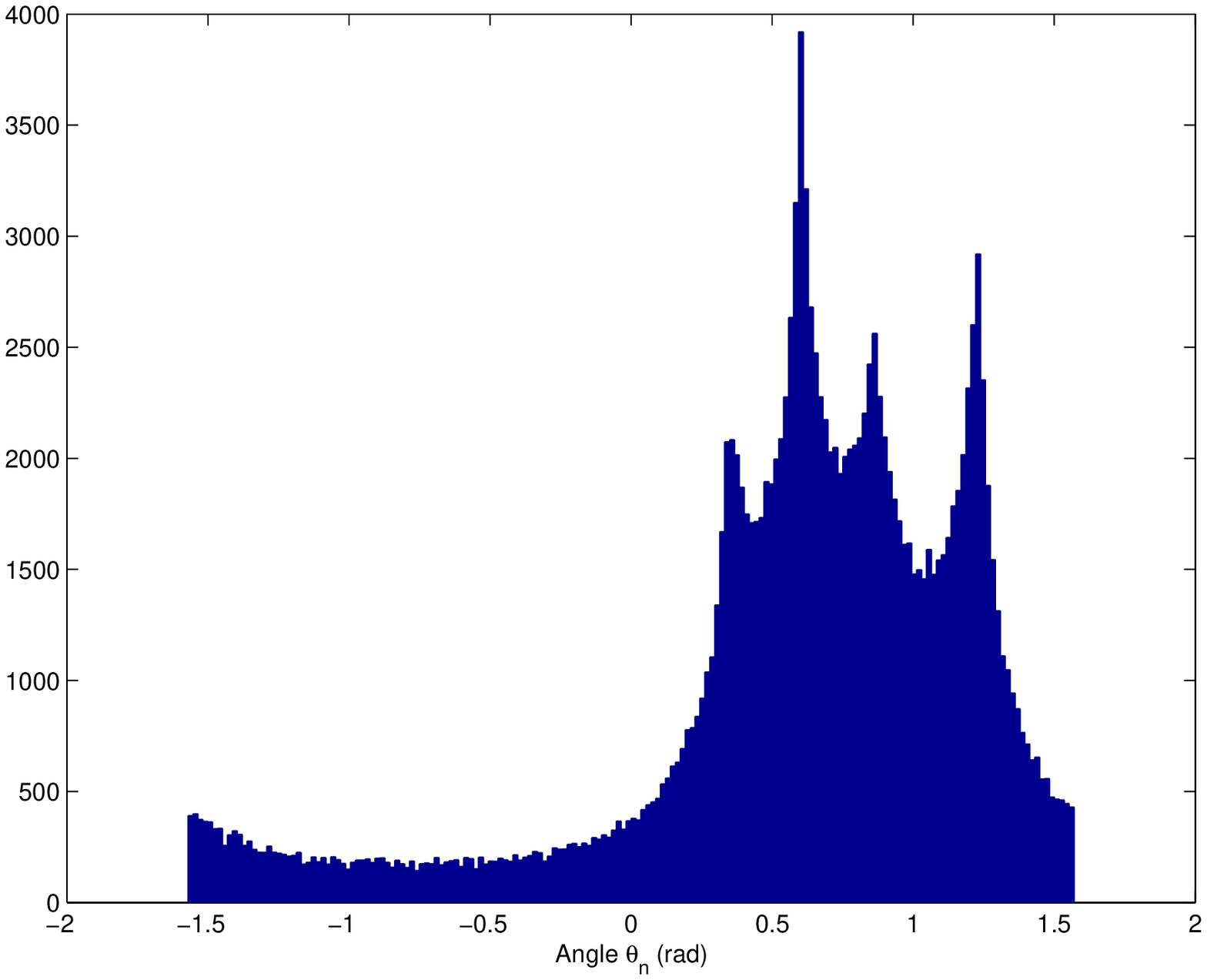}}
\subfigure[Reduced Histogram \cite{Mitianoudis07f} ]{\includegraphics[width=1.7in]{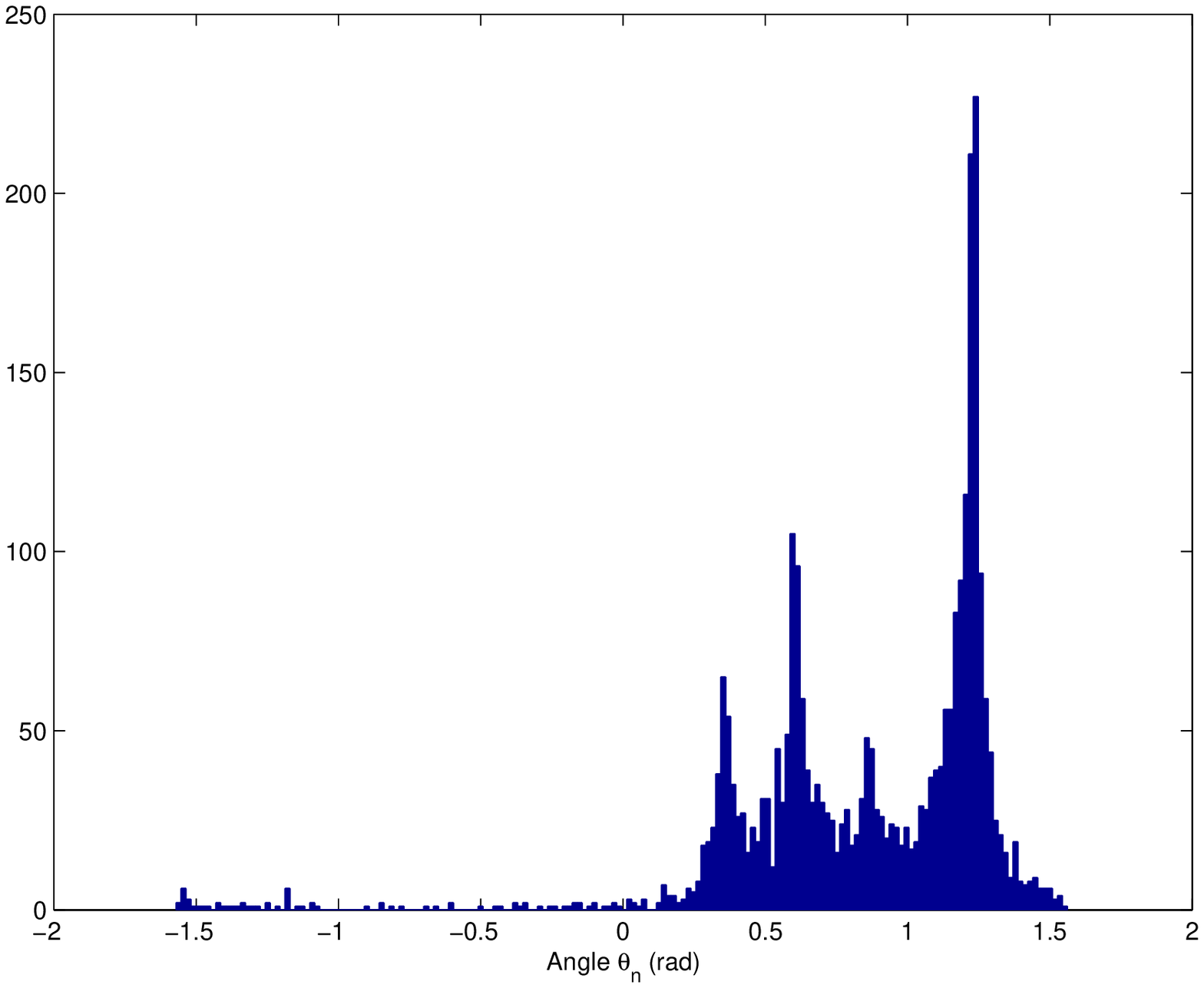}}
\subfigure[Reduced Histogram \cite{Arberet10} ]{\includegraphics[width=1.7in]{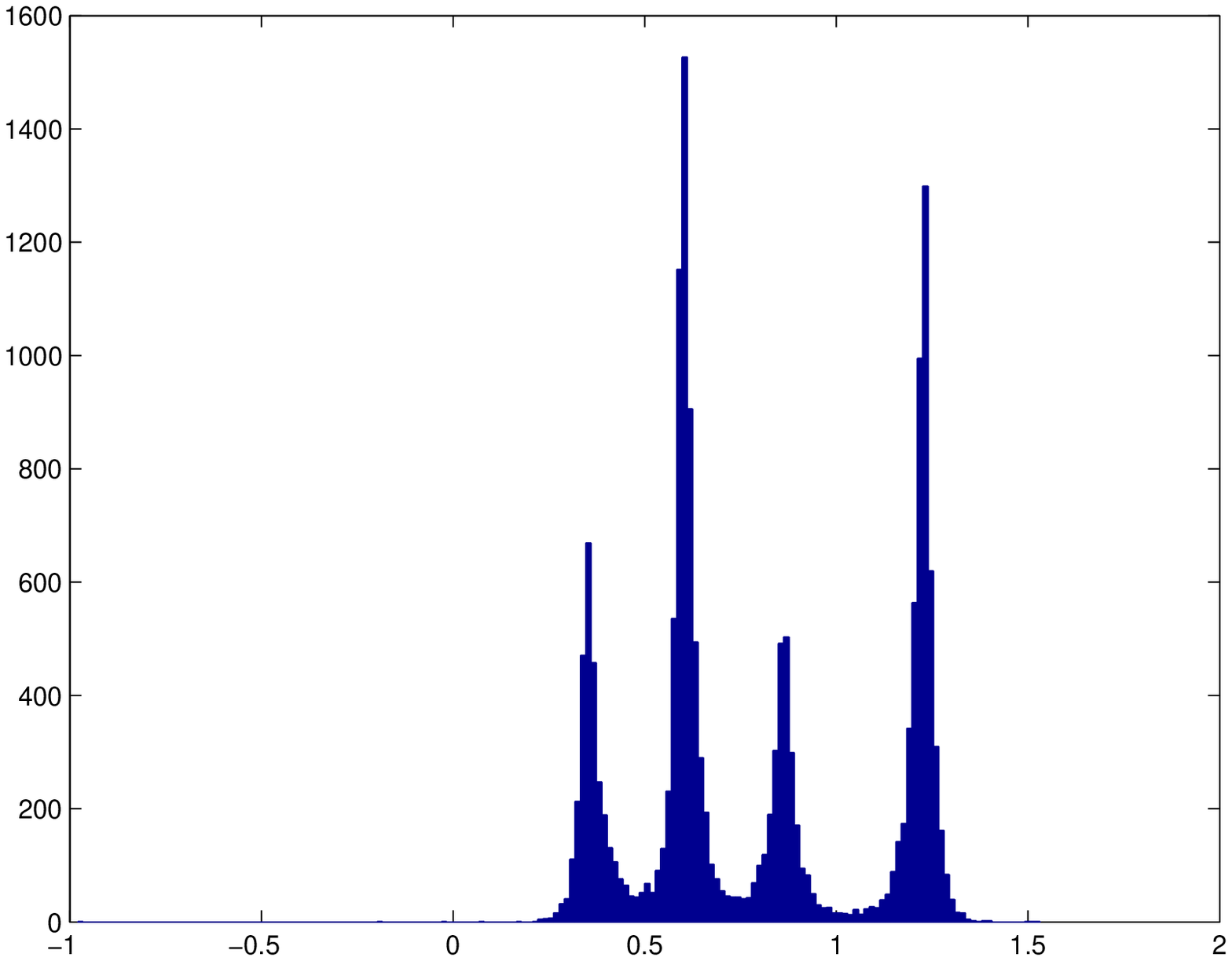}}\\
\caption{Comparison between the two mechanisms of sparsifying the mixture data in the transform domain for a $2\times 4$ scenario. The simplified Arberet et al \cite{Arberet10} method offers more distinct concentrations along the source mixing directions.}
\label{fig1}
\end{figure*}

In \cite{Mitianoudis12}, Mitianoudis introduced a multidimensional Directional Laplacian Density (DLD) model as a closed form solution to the problem of modelling directional sparse data. Mixtures of DLD (MDLD) were also used to address the general $K\times L$ underdetermined source separation problem, with promising results. In \cite{Karavasilis11}, a variant of the common Gaussian Mixture Model (GMM) was proposed in order to enforce weights on the data points contributing to the estimation of the GMM's parameter, according to the distance of each data point to each Gaussian's centre. The technique was coined {\em weighted GMM}. As in every underdetermined separation effort, source sparsity is usually a prerequisite in order to facilitate separation. In \cite{Mitianoudis12}, the Modified Discrete Cosine Transform (MDCT) was used mainly due to its real-valued output.  In this paper, we replace the MDCT framework with the sparse STFT framework similar to the one proposed by Arberet et al~\cite{Arberet10}. Now, a more careful selection of time-frequency points is being made in order to sparsify the signal representation and the use of a weight parameter to estimate the contribution of each point in the DLD mixtures aid the separation effort. In addition, we extend the work of\cite{Mitianoudis12} by deriving weighted Mixtures of Directional Laplacians, in order to emphasize the contribution of points that are closer to the cluster centres.
\section{Sparse STFT Time-Frequency Representation}
 In the time domain representation, many of the mixture's characteristics are not visible and the separation problem is thus more difficult to solve. The solution is to get a sparser representation of the input data by applying the Short Time Fourier Transform (STFT) $X_k(t,f)$ on each channel $x_k(n)$ of the mixture. As a result, the mixing model is approximately written in a complex matrix form in the time-frequency domain as $\textbf{X}(t,f)=\textbf{A}(f)\textbf{S}(t,f)$, where $t$ is the time frame index and $f$ represents the normalized frequency, $\textbf{X}(t,f)=[X_1(t,f),...,X_K(t,f)]^T$ and $\textbf{S}(t,f)=[S_1(t,f),...,S_L(t,f)]^T$. 

In \cite{Arberet10}, Arberet et al made a similar assumption to Yilmaz and Rickard \cite{Yilmaz04}, that for every source there is at least one time-frequency area, where each source dominates over others. This implies that there exist several time-frequency points $(t,f)$, where the $l$-th source is only present. In \cite{Arberet10}, Arberet et al proposed a method to identify those time-frequency regions, which is slightly simplified here. For each time-frequency point $(t,f)$, we consider a time-frequency region $\Omega_{t,f}$ ``in the neighbourhood'' of this point, i.e. a window of size $Q\times Q $ centred around each time-frequency point $(t,f)$. Each region yields a complex-valued local scatter plot $\textbf{X}(\Omega)$ and thus by employing Principal Component Analysis (PCA) on $\textbf{X}(\Omega)$, one can derive a \emph{local confidence measure} ${T(\Omega_{t,f})}$, which is greater in the case of a single source present in the neighbourhood $\Omega_{t,f}$ and smaller in the opposite case of no sources or more than one source. To simplify the method in \cite{Arberet10}, one can observe that since the mixing is instantaneous, then $\mathbf{A}(f)\in\Re^L$. Thus, instead of calculating the complex product  $\textbf{X}(\Omega)\textbf{X}^H(\Omega)$, one can concatenate the real and imaginary parts of $\textbf{X}(\Omega)$, producing the augmented matrix $\textbf{X}_{aug}(\Omega)=[\textrm{Re}\{\textbf{X}(\Omega)\} ; \textrm{Im}\{\textbf{X}(\Omega)\}] $ and perform the previous task using the real covariance matrix of $\textbf{X}_{aug}(\Omega)$. This is valid since the real-valued instantaneous mixing will be equally applied to both the real and imaginary parts of $\textbf{X}(\Omega)$.  Using the real-valued PCA, we can obtain a principal direction as a unit vector $\hat{\textbf{u}}(\Omega)$ and the real-valued positive eigenvalues in decreasing order $\lambda_1(\Omega)\geq\lambda_2(\Omega)\geq ...\lambda_K(\Omega)$ of the $K\times K$  positive definite covariance matrix $C_X=\textbf{X}_{aug}(\Omega)\textbf{X}_{aug}^T(\Omega)$. Therefore, a confidence measure can be defined as:
\begin{equation}
{T}(\Omega):={\lambda}_1(\Omega)/\frac{1}{K-1}\sum_{k=2}^K{\lambda}_k(\Omega)
\end{equation}
 Once the local confidence measure ${T(\Omega_{t,f})}$ is above a threshold $d$, then this neighboroughood contains only one source. Thus, if we use points from similar single-source areas, we can train a clustering algorithm in order to perform separation. This is more efficient than using all available time-frequency points, since in the new reduced dataset, the selected points are placed more dominantly along the source directions. 

This mechanism is a more intelligent method to remove points that do not contribute to the source's mixing directivity. A simple thresholding of $r(n)=||\textbf{x}(n)||_2$ was proposed in \cite{Mitianoudis07f} to remove points close to $r(n)<\textrm{Threshold}$, which served as a heuristic, but not intelligent, method to achieve the previous task. In Fig. \ref{fig1}, one can see the differences between the two sparsification mechanisms and verify that the above described method offers smoother and more emphasized concentrations along the columns of the mixing matrix.

\section{Mixtures of Weighted Directional Laplacian Distributions}
The Generalised $D$-dim Directional Laplacian Distribution (DLD) is given by the following equation \cite{Mitianoudis12}:
\begin{equation}
p(\mathbf{x})=c_D(k)e^{-k\sqrt{1-(\mathbf{m}^T\mathbf{x})^2}}\quad, \forall\textrm{ }||\mathbf{x}||\in \mathcal{S}^{D-1}
\label{MDDLD}
\end{equation}
where $\mathcal{S}^{D-1}$ is the $D$ dimensional unit hypersphere, $\mathbf{m}$ defines the mean, $k\geq 0$ defines the width (``approximate variance'') of the distribution. Now, $c_D(k)=\frac{\Gamma(\frac{D-1}{2})}{\pi^{\frac{D+1}{2}} I_{D-2}(k)}$, $I_{D}(k)=\frac{1}{\pi}\int_{0}^{\pi} e^{-k\sin\theta}\sin^{D}\theta d\theta$ is the normalising coefficient and $\Gamma(\cdot)$ represents the Gamma function. We can use {\em Weighted Mixtures of Generalised Directional Laplacians} (WMDLD) in order to model multiple concentrations of directional ``heavy-tailed signals'', such as those encountered in underdetermined source separation scenarios.
\begin{equation}
p(\mathbf{x}_n)=\sum_{i=1}^R w_{n,i}a_i c_D(k_i)  e^{-k_i\sqrt{1-(\mathbf{m}_i^T\mathbf{x})^2}}, \forall\textrm{ }||\mathbf{x}_n||\in \mathcal{S}^{D-1}
\label{MMDDLD}
\end{equation}
where $a_i$ denotes the weight of each distribution in the mixture, $R$ the number of DLDs used in the mixture and $\mathbf{m}_i$, $k_i$ denote the mean and the approximate variance of each distribution.  DLD mixtures can be commonly trained using the Expectation-Maximisation (EM) algorithm\cite{Mitianoudis12}. The innovation in this paper is to introduce a weight parameter that can hinder the uniform influence of a data point to every DLD in the mixture during the learning process, but instead increase the influence to those DLDs that describe this data point more accurately. This is achieved by introducing the weight parameter $w_{n,i}$, which gets a higher value if the data point $\mathbf{x}_n$ is closer to the DLD's mean vector $\mathbf{m}_i$ and the opposite. First, we need to introduce a suitable distance function for directional data, such as the following:
\begin{equation}
D_l(\mathbf{x}_n,\mathbf{m}_i)=\sqrt{1-(\mathbf{m}_i^T\mathbf{x}_n)^2}
\label{DLWEIGHT}
\end{equation}
The proposed distance function is monotonic and emphasises the contribution of the points closer to each distribution center $\mathbf{m}_i$. In addition, it is similar to a L1-norm, rather than a L2-norm distance (see \cite{Mitianoudis12}), thus it fits sparse data more appropriately. The proposed weight parameter $w_{n,i}$ should use $D_l(\mathbf{x}_n,\mathbf{m}_i)$ and should attribute greater values for those points where $D_l$ is small and vice versa. Thus, the following mapping gives the desired $w_{n,i}$ properties:
\begin{equation}
w_{n,i}=\frac{1}{2}(1-D_l(\mathbf{x}_n,\mathbf{m}_i))
\label{WEIGHT}
\end{equation}
Deriving the updates of the EM is straighforward, following the derivations in \cite{Mitianoudis12} and \cite{Karavasilis11}. The updates for $p(i|\mathbf{x}_n)$ and $\alpha_i$ are provided below:
\begin{equation}
p(i|\mathbf{x}_n)\leftarrow \frac{a_i c_D(k_i) e^{-k_i\sqrt{1-(\mathbf{m}_i^T\mathbf{x}_n)^2}}}{\sum_{i=1}^Ka_i c_D(k_i) e^{-k_i\sqrt{1-(\mathbf{m}_i^T\mathbf{x}_n)^2}}}
\end{equation}
\begin{equation}
a_i\leftarrow \frac{1}{N}\sum_{n=1}^Nw_{n,i}p(i|\mathbf{x}_n)
\end{equation}
Similarly, the updates for $\mathbf{m}_i$ and $k_i$ are given below:
\begin{equation}
\mathbf{m}_i^+\leftarrow \mathbf{m}_i +\eta\sum_{n=1}^N k_i\frac{\mathbf{m}_i^T\mathbf{x}_n}{\sqrt{1-(\mathbf{m}_i^T\mathbf{x}_n)^2}}\mathbf{x}_nw_{n,i}p(i|\mathbf{x}_n)
\label{estimM3}
\end{equation}
\begin{equation}
\mathbf{m}_i^+\leftarrow \mathbf{m}_i^+/||\mathbf{m}_i^+||
\end{equation}
To estimate $k_i$, we solve the equation ${\partial I}/{\partial k_i}=0$ numerically (see \cite{Mitianoudis12}). The equation yields:
\begin{equation}
\frac{I_{D-1}(k_i)}{I_{D-2}(k_i)}=\frac{\sum_{n=1}^{N}\sqrt{1-(\mathbf{m}_i^T\mathbf{x}_n)^2}w_{n,i}p(i|\mathbf{x}_n)}{\sum_{n=1}^{N}w_{n,i}p(i|\mathbf{x}_n)}
\label{estimK3}
\end{equation}

Once the WMDLD model is trained on the reduced data, it can be employed to separate the complete dataset.

\section{Source Separation scheme}
Once the Weighted Mixtures of DLD are fitted to the multichannel directional data, separation on the complete dataset can be performed by ``hard-thresholding'' for the 1-dim case in a similar manner to~\cite{Mitianoudis12}. That is to say, we estimate the intersection points between the estimated DLDs, which determine the hard thresholds that can attribute input data points to the different sound sources. For the $D$-dim case, a ``soft-thresholding'' technique was employed in ~\cite{Mitianoudis12}, since the calculation of intersection planes in the $D$-dim case is not straightforward. Here, we propose another ``hard-thresholding'' (``winner-takes-all'') clustering strategy. We calculate the distance between each data point and the estimated DLD centres $m_i$ using the directional distance of (\ref{DLWEIGHT}). Evidently, each data point is attributed to the DLD with the minimum distance from its centre.

Having attributed the points $\mathbf{x}_n$ to the $L$ sources, the next step is to reconstruct the sources. Let $S_i\sqsubseteq N$ represent the data point indices (samples) that have been attributed to the $i^{th}$ source and $\mathbf{m}_i$ the corresponding mean vector, i.e. the corresponding column of the mixing matrix. We initialise $u_i(n)=0 , \forall$ $n=1,\dots,N$ and $i=1,\dots,L$. The source reconstruction is performed by substituting:
\begin{equation}
u_i(S_i)=\mathbf{m}_i^T\mathbf{x}_{S_i}\qquad \forall\textrm{ }i=1,\dots,L
\end{equation}
The source signals $u_i$ are then moved back to the time-domain using the inverse STFT. 

\section{Experiments}
In this section, we evaluate the proposed WMDLD algorithm  for audio source separation. We will use the MoWL algorithm~\cite{Mitianoudis07c} the ``GaussSep'' algorithm~\cite{Vincent09} and the original MDLD~\cite{Mitianoudis12} for comparison. After fitting the MDLD with the proposed EM algorithm, separation will be performed using hard or soft thresholding, as described earlier. For quantitative evaluation, we calculate the {\em Signal-to-Distortion Ratio} (SDR), the {\em Signal-to-Interference Ratio} (SIR) and the {\em Signal-to-Artifact Ratio} from the BSS$\_$EVAL Toolbox v.3~\cite{BSSeval}. The input signals for the MDLD and MoWL approaches are sparsified using the {\em Modified Discrete Cosine Transformation} (MDCT). The frame length for the MDCT analysis is set to $32$ msec for the speech signals and $128$ msec for the music signals sampled at $16$ KHz, and to $46.4$ msec for the music signals at $44.1$ KHz. We initialise the parameters of the MoWL and MDLD as follows: $\alpha_i=1/N$ and $c_i=0.001$,  $T=[-1,0,1]$ (for MoWL only) and $k_i=15$ (for the DLD only). The centres $m_i$ were initialised in either case using the Directional {\em K-means} step, as described in \cite{Mitianoudis12}.  We used the ``GaussSep'' algorithm, as publicly available by the authors\footnote{MATLAB code for the ``GaussSep'' algorithm is available from {\tt http://www.irisa.fr/metiss/members/evincent/software}.}. For the estimation of the mixing matrix, we used Arberet et al's~\cite{Arberet10} DEMIX algorithm\footnote{MATLAB code for the ``DEMIX'' algorithm is available from {\tt http://infoscience.epfl.ch/record/165878/files/}.}, as suggested in \cite{Vincent09}, combined with the weighted DLDs. The number of sources in the mixture was also provided to the DEMIX algorithm, as it was provided to all other algorithms. The ``GaussSep'' algorithm operates in the STFT domain, where we used the same frame length with the other approaches and  a time-frequency neighbourhood size of $5$ for speech sources and $15$ for music sources.  For the proposed WMDLD, we used an STFT with the same frame length as with the other approaches, similar settings for the mixture model as with the MDLD. For the sparse STFT framework, we used a window size of $Q=2$ for speech sources and $Q=3$ for music sources and a threshold for selecting the appropriate $t-f$ points between $T(\Omega_{t,f})=300-350$.

The algorithms were tested with the {\em Groove}, {\em Latino1} and {\em Latino2} datasets~\cite{BASS-dB} ($44.1$ KHz sampling frequency). Signal Separation Evaluation Campaigns SiSEC2008~\cite{SiSEC2008} and SiSEC2010~\cite{SiSEC2010} provided many other test signals. We used two audio instantaneous mixtures the ``Dev2WDrums'' and  ``Dev1WDrums'' sets (3 instruments at 16KHz) and two speech instantaneous mixtures the ``Dev2Male3'' and ``Dev2Female3'' sets (4 closely located sources at 16 KHz).

In order to test multichannel separation, we used the Dev3Female3 set from SiSEC2011~\cite{SiSEC2011}, a $3\times 5$ (3 mixtures - 5 sources) and a $4\times 8$ (4 mixtures - 8 sources) scenario with random male and female voices. For the $3\times 5$ example, we mixed 5 speech sources around the angles $\theta_1= [0^o,-87^o,-60^o,0^o,45^o]$ and $\theta_2 = [85^o,0^o,-60^o,0^o,45^o]$.  For the $4\times 8$ example, eight audio sources were mixed around the angles: $\theta_1= [-75^o,-30^o,0^o,50^o,10^o,80^o, -45^o, 0^o]$,   $\theta_2= [ 70^o,30^o,-20^o,50^o,-70^o,0^o,15^o,-70^o]$ and $\theta_3= [ 80^o,20^o,10^o,-50^o, 0^o,-10^o,-25^o, -35^o ]$. Readers can visit the following url\footnote{\label{foot1} {\tt http://utopia.duth.gr/nmitiano/mdld.htm}} and listen to the separation results. 

In Table \ref{Table5}, we can see the estimated values of SDR, SIR and SAR for each of the methods we described above. We have also averaged the scores for all sources at each experiment. The values of the proposed WMDLD approach show improvement in all cases compared to the MDLD method, probably due to the confidence measure with which our method selects the most significant time-frequency points ($t,f$) and the improved EM training due to the importance weights introduced to the input  data points. In comparison to the ``GaussSep'' method, WMDLD is better in terms of the SIR index, but is still falling behind in terms of the SDR and SAR indexes, meaning that our method removes more interference from other sources in the mixture, but there are still more artifacts comparing to ``GaussSep'' method.

The results of the $K>2$ case are shown in Table \ref{Table7}. Similar to the $K=2$ case, the proposed WMDLD features higher performance than the MDLD method in all terms. It is important to notice that although the ``GaussSep'' method continues to show better values than the WMDLD approach in terms of SDR and SAR in $K=3$ case, in the $K=4$ case ``GaussSep'' fails to separate the sources when WMDLD, in contrast, separates the 8 sources. The main difference between ``GaussSep'' and WMDLD is that WMDLD is stricter in terms of separation, thus eliminating more the crosstalk between the sources compared to ``GaussSep''. This justifies its higher SIR value. However, since the separation is stricter, thus more points are being uniquely clustered to a single source, which will give rise to more reconstruction artifacts. This justifies its slightly lower SDR and SAR values. Nevertheless, WMDLD is a faster approach which can perform separation to than $K>3$ sensor signals.

\begin{table*}[htb]
\caption{The proposed WMDLD approach is compared ($K=2$) in terms of SDR (dB), SIR (dB) and SAR(dB) with MDLD, GaussSep (GS) and MoWL approach. The measurements are averaged for all sources of each experiment.} 
\begin{center}
{
\begin{tabular}[width=0.5\textwidth]{|l||c|c|c|c||c|c|c|c||c|c|c|c|}\hline
&\multicolumn{4}{c||}{\bf{SDR} (dB)}&\multicolumn{4}{c||}{\bf{SIR} (dB)}&\multicolumn{4}{c|}{\bf{SAR} (dB)}\\\hline
&{\small WMDLD}&{\small MDLD}&{\small GS}&{\small MoWL}&{\small WMDLD}&{\small MDLD}&{\small GS}&{\small MoWL}&{\small WMDLD}&{\small MDLD}&{\small GS}&{\small MoWL}\\\hline\hline
{\ssmall Latino1}&7.62&6.38&5.51&5.72&16.84& 18.63&	8.96&	18.59&8.52&6.93&9.20&6.26\\\hline
{\ssmall Latino2}&4.98& 3.21&4.71&2.10&12.62&  11.50&8.87&	11.28&7.17& 4.95&	9.20&	3.85\\\hline
{\ssmall Groove}&2.34&0.22&0.39&-0.43&11.89&9.48&3.62&9.60&4.04&2.12&7.37&1.00\\\hline
{\ssmall Dev2Male4}&4.68&3.04&6.22&	2.11&	14.46& 13.69&	12.14&	13.30&	5.76& 4.10&8.04&3.33\\\hline
{\ssmall Dev2Female4}&6.09& 4.68&5.70&3.86&	16.75&  15.28&	11.45&	16.58&	6.90& 5.41&	7.51&4.61\\\hline
{\ssmall Dev2WDrums}&10.33&9.59&16.57&10.16& 19.61&19.77&23.83&19.98&11.26&10.55&	17.68&10.54\\\hline
{\ssmall Dev1WDrums}&8.94&4.96&16.54&3.81&17.71& 13.88&20.94&	12.38&	9.97&6.37&	19.30&	5.20\\\hline\hline
\it{Average}&6.43&4.58	&\bf{7.96}	&3.91&	\bf{15.70}& 14.61&	12.83	&13.82&	7.66& 5.78&\bf{11.19}	&4.97\\\hline
\end{tabular}
}
\end{center}
\label{Table5}
\end{table*}

\begin{table*}[htb]
\caption{The proposed WMDLD approach is compared for source estimation performance ($K=3,4$) in terms of SDR (dB), SIR (dB) and SAR(dB) with the MDLD and the GaussSep (GS) approach. The measurements are averaged for all sources of each experiment.} 
\begin{center}
{
\begin{tabular}[width=0.0.5\textwidth]{|l||c|c|c||c|c|c||c|c|c|}\hline
&\multicolumn{3}{c||}{\bf{SDR} (dB)}&\multicolumn{3}{c||}{\bf{SIR} (dB)}&\multicolumn{3}{c|}{\bf{SAR} (dB)}\\\hline
&WMDLD&MDLD&GS&WMDLD&MDLD&GS&WMDLD&MDLD&GS\\\hline\hline
Dev3Female4&11.77&6.02&16.93&22.26&23.84&22.43&12.23&6.17&18.40\\\hline
Example $3\times5$ &8.41&3.91&9.94&17.58&17.92&15.21&9.10&4.17&11.68\\\hline
Example $4\times8$ &5.29&2.24&-18.63&13.72&16.4&-17.58&6.23&2.52&9.39\\\hline

\end{tabular}
}
\end{center}
\label{Table7}
\end{table*}

\section{Conclusions}

In this paper, we extended our previous work on MDLD, replacing the MDCT with a STFT framework, where the existence of a single dominant source in time-frequency neighbourhood is ensured in order to sparsify the data. The next improvement is the introduction of importance weights for each data point during EM adaptation, which improves the training accuracy of the model and thus its separation performance. Finally, a simpler and more effective hard-thresholding strategy is proposed to perform $p$-dim source separation. For future work, we will be looking into extending this framework for convolutive mixtures.

\bibliographystyle{IEEEbib}
\bibliography{ica}

\end{document}